\documentclass[useAMS,usenatbib]{mn2e}

\usepackage{lscape}

\usepackage[usenames,dvips]{color}
\usepackage{graphicx}
\usepackage{epsfig}

\title[Three-dimensional Two-Layer Outer Gap Model]{Three-dimensional Two-Layer Outer Gap Model: $Fermi$ Energy Dependent Light Curves of the Vela Pulsar}

\author[Y. Wang, J Takata $\&$ K.S. Cheng]{Y. Wang, J. Takata and K.S. Cheng \thanks{E-mail:
yuwang@hku.hk (YW); takata@hku.hk (JT); hrspksc@hkucc.hku.hk (KSC)}
\\
Department of Physics, University of Hong Kong, Pokfulam Road, Hong Kong}

\begin{document}

\pagerange{\pageref{firstpage}--\pageref{lastpage}} \pubyear{2010}

\maketitle

\label{firstpage}

\begin{abstract}
We extend  the two-dimensional two-layer outer gap model to a three-dimensional geometry and use it to study the {{}high-energy emission} of the Vela pulsar.
In this model,  the outer gap is divided into two parts, i.e. the
 main acceleration region on the top of last-open field lines and the screening
region around the upper boundary of the gap. In the main acceleration region,
 the charge density is much lower than the Goldreich-Julian charge density
and the charged particles are accelerated by the electric field along
the magnetic field to emit multi-GeV photons. In the screening region, the charge density is
larger than the Goldreich-Julian value to close the gap and particles in this region are responsible for multi-100MeV photon emission.
 We apply this three dimensional two-layer model to the Vela pulsar and compare the model
light curves, the phase-averaged spectrum and the phase-resolved spectra with the recent $Fermi$ observations, which also reveals the existence of the  third peak between two main peaks.
The phase position of the third peak moves with
the photon energy, which cannot be explained by the geometry of magnetic field structure and
 the caustic effects of the photon propagation.
We suggest that the existence of the
third peak and its energy dependent movement {{}results from} the azimuthal structure
of the outer gap.
\end{abstract}


\section{Introduction}
{{}The Vela pulsar is one of the few $\gamma$-ray pulsars, whose $\gamma$-ray
radiations are so intense that could be detected before the launch of $Fermi$ $\gamma$-ray telescopes.}
 For this reason, the observed spectra and the pulse profiles
 of the Vela pulsar have been studied extensively
before the {{}launch} of $Fermi$ (Romani, 1996; Dyks, Harding \& Rudak, 2004; Takata, Chang \& Shibata
2008).  {{}Its $\gamma$-ray spectrum between 100MeV to several GeV is difficult to be explained in terms of simple curvature radiation.} Takata, Chang \& Shibata (2008) proposed that the component around 100MeV of the Vela pulsar, as well as Geminga (Takata \& Chang, 2009), is due to the synchrotron radiation of the incoming particles in the outer gap. By solving the evolution of the Lorentz factor and the pitch angle in the two-dimensional electrodynamics model (Takata, Shibata \& Hirotani, 2004), they found that the synchrotron radiation of the incoming particles dominates the radiation around 10 MeV to 100MeV and the curvature radiation of the outgoing particles dominates the radiation around 1GeV.

 Zhang \& Cheng (1997) proposed a self-sustained thick outer gap model of $\gamma$-ray emission from the rotation-powered pulsar. {{}They pointed out that the primary $e^{\pm}$ pairs in the steady state have a power-law distribution and used this distribution to calculate the spectra of some $\gamma$-ray pulsars}, including Geminga-like and Vela-like pulsars. Also based on this distribution of primary pairs, Zhang \& Cheng (2001) applied a three-dimensional pulsar magnetosphere model to explain the high-energy emission from the Geminga pulsar with a thick outer gap. In this calculation, the high-energy $\gamma$-rays are produced by the accelerated particles with a power-law energy distribution via curvature radiation inside the outer gap.

After the launch of $Fermi$ LAT, the bottleneck of the research of high energy radiation from pulsar is broken.  Just in one year, this high-quality $\gamma$-ray telescope has measured 46 $\gamma$-ray pulsars (Abdo et al. 2010a).
We developed a two-layer model (Wang et al., 2010) to study the phase averaged
spectra of the mature pulsars, including the Vela pulsar, in the first catalogue of $\gamma$-ray pulsars of $Fermi$. In this simple two-dimensional model, the outer gap extending outwards from the null charge surface is considered as a superposition of a main acceleration region and a screening region. In the main acceleration region, which is close to vacuum, the electric field is strong, and  the photons generated in this region dominate the radiation of around GeV range.  In the screening region, the large numbers of the pairs produced by the
pair-creation process stop the growth of the acceleration region and they  produce the photons with around 100MeV range. Although
 this two-dimensional two-layer model could  explain the phase-average $\gamma$-ray spectra of the mature pulsars,
it  cannot be used to study  the phase-resolved spectra and the light curves.
Furthermore the high quality data of $Fermi$ has clearly shown the existence of a third peak (bump) at the trailing part of the first peak, which was explained as an effect of geometry of the magnetic field (Dyks \& Rudak, 2003; Fang \& Zhang, 2010). However, the observed result provided by $Fermi$ (Abdo et al, 2010b) shows that, as the energy increases, this third peak shifts towards the second peak. This indicates that unlike those two main peaks this third peak does not result from the caustic effect of photon propagation and/or the structure of the magnetic field.

 The high-energy radiation processes of pulsars have been studied based on polar cap models
(Ruderman \& Sutherland 1975; Daugherty \& Harding, 1982),
 the slot gap models (Arons 1983; Harding, Usov \& Muslimov 2005;
 Harding et al. 2008) and the outer gap  models (Cheng, Ho \& Ruderman 1986a,b; Romani 1996; Hirotani 2008; Takata \& Chang 2009).  The different
acceleration  models have predicted different properties of the
$\gamma$-ray emission properties from the pulsar magnetosphere.
To discriminate among these existed emission models, it is better to compare the predicted phase resolved spectra with the observed ones, which contains the most detailed information about the structure of pulsar magnetosphere and the acceleration mechanism.

In this paper, we extend our study of two-layer model from two-dimensional to three-dimensional, which includes the effects of
the inclination angle, viewing angle and the azimuthal structure of the gap. 
{{}This three-dimensional model can provide the phase-resolved spectra and the energy-dependent light curves that cannot be given by the two-dimensional one.}
We will use it to study  the observed energy dependent light curves of the Vela pulsar measured  by $Fermi$ LAT (Abdo et al, 2010b).

In section~2, we describe our  theoretical model, which contains the electrodynamics of the outer gap, the emission geometry and the method to calculate the
 spectra and the light curves. In section 3, we will use our  three-dimensional model to explain the energy dependent light curves of the Vela pulsar and discuss the conditions of the existence of and the movement of the third peak as well as other observed spectral properties. We gives a brief summary in  section 4.

\section{Theoretical Model}

\subsection{Three dimensional gap structure}
\begin{figure}
\centerline{\psfig{figure=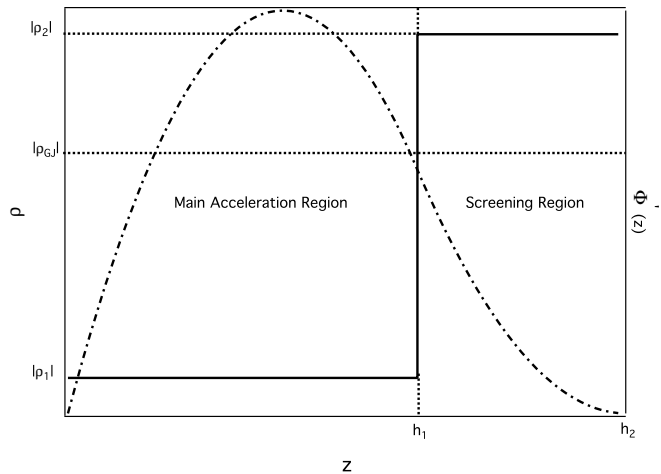, width=9.2cm, clip=}}
\caption{The simplified distribution of the charge density (solid line) and the corresponding accelerating potential (dot-dashed line) of the two-layer outer gap.}
\label{struct}
\end{figure}

We discuss the $\gamma$-ray emissions from the outer gap accelerator in the
pulsar magnetosphere, and  assume
that a strong acceleration region is extending
above  the last-open filed lines and between the
null charge surface of the Goldreich-Julian charge density
and the light cylinder. Although the dynamical studies
 (e.g. Takata, Shibata \& Hirotani, 2004) predict that the inner boundary of the active gap
 can extend inwards from the null charge surface, most of the
power of the gap are released beyond the null charge surface. Because the pairs are
always created on convex side of the magnetic field lines, the charge number
density increases as the height measured from the last-open field lines.
In the trans-field direction of the magnetic field, therefore, as shown in Figure~\ref{struct}, the outer gap
 can be divided into two parts:
1) the main acceleration region at the lower part
of the outer gap, where the charge density is assumed to be $\sim$ 10~\% of the
Goldreich-Julian value and a strong electric field is accelerating
the particles to emit GeV photons via the curvature process,
 and 2) the screening region around the upper boundary, where
 the growth of the main acceleration region in the trans-field direction is stopped
 by the pair-creation process.
 {{}Such two-layer structure, in the direction along the field line, starts from a place between the inner boundary and the null charge surface, and ends at the light cylinder. Inside the inner boundary and outside the light cylinder, there is no accelerating electric field; from the null charge surface to the inner boundary, the accelerating electric field decreases rapidly to zero.}

 Because the number density of the pairs made by the pair-creation process increases
exponentially  in the trans-field direction
(Cheng, Ho \& Ruderman 1986a, b), we use
 a simple step function to approximate the distribution of the charge density
 in the trans-field direction in the poloidal plane {(the plane where the field lines have same polar angle $\phi_p$)} as
 \begin{equation}
\rho(x, z, \phi_p)=\left\{
   \begin{array}{ccc}
            \rho_1(x, \phi_p), &$if$& 0\leq{}z\leq{}h_1(x, \phi_p)\\
            \rho_2(x, \phi_p), &$if$& h_1(x, \phi_p)<z\leq{}h_2(x, \phi_p)
   \end{array},\right.
\end{equation}
where $x$, $z$ and $\phi_p$ represent coordinates along the magnetic field line,
the height measured from the last-open field line, and the azimuthal direction.
In addition, $h_1$ and $h_2$ represent the thickness of the main-accretion
region and the total gap thickness, respectively.
As illustrated in Figure~\ref{struct}, in the $z$-direction, the main acceleration region
 spans from $z=0$ to $z=h_1$, where the charge density is substantially smaller than the
Goldreich-Julian charge density (Goldreich \& Julian 1969), and the screening
region is between $h_1\le{}z\le{}h_2$, the charge density in it is much higher than
 $\rho_{GJ}$ in order to close the gap. For each poloidal plane ($\phi_p$=constant),
 $h_1(x,\phi_p)$ as well as $h_2(x,\phi_p)$
is defined with a given magnetic field line in the constant $\phi$-plane
so that the ratio $h_1(x,\phi_p)/h_2(x,\phi_p)$ is nearly constant along
the field lines.

Even though the structure of the gap is simplified as described above,
the three-dimensional
 Poisson equation with the magnetic field structure
is still difficult to be solved in comparing with
 the case of the two-dimensional model discussed  in Wang et al.(2010).
Instead of solving the real three-dimensional Poisson equation, we adopt
 a simplified Poisson equation. We calculate  the
electric field structure at a fixed poloidal plane, that is, at each azimuthal
coordinate by solving the two-dimensional Poisson equation. This approximation can be
justified if the derivation of the potential field in the
azimuthal direction is much smaller than that in poloidal plane.
This procedure
may be applicable for the thin outer gap accelerator, whose
thickness in the poloidal plane is much smaller than the width in the azimuthal
direction. The detail procedure to obtain the three-dimensional structure
is described as follows.

In a fixed poloidal plane, the bottom and top boundaries of the outer gap are the last-open field lines and a surface layer with separation $h_2$
made by  the magnetic field lines respectively. We
divide
the polar cap into $N_B$ equal divisions in the azimuthal direction.
If the gap size of the azimuthal direction is much larger
than the gap thickness in the poloidal plane, the gap structure of each  slice with a fixed
azimuthal angle $\phi_p$ may be approximately
represented by  the two-dimensional situation shown in Wang et al. (2010).

The accelerating electric field in each slice
of the gap is  solved by the
method described in the previous two-dimensional model(Wang et al., 2010).
 The potential in the gap is described by  the Poisson equation,
 $\bigtriangledown^2\Phi=-4\pi\rho$.
This potential
can be written as $\Phi=\Phi'+\Phi_{0}$, where $\Phi_0$ is the co-rotating potential, which satisfies $\bigtriangledown^2\Phi_0=-4\pi\rho_{GJ}$, and the potential $\Phi\rq{}$  represents the deviation from co-rotating  potential. This deviation generates the accelerating electric field. {{}By assuming the derivation of the potential field in the azimuthal direction is small enough to be neglected compared with that in the poloidal plane}, we express the Poisson equation of $\Phi'$ as
\begin{equation}
\left(\frac{\partial^2}{\partial x^2}
+\frac{\partial^2}{\partial z^2}\right)\Phi'=-4\pi(\rho-\rho_{GJ}).
\label{Poisson0_3}
\end{equation}
 Because our two-dimensional study predicts that the outer gap of the Vela pulsar is thin, the variation of the Goldreich-Julian charge
density in the trans-field direction (that is $z$-direction) may be able to
be neglected so that  the $\rho_{GJ}$ can be assumed to be constant in $z$-direction.

 Moreover, we assume that  the
 derivative  of the potential field in the trans-field (z) direction
is larger than that along (x-direction) the  magnetic field line.
This assumption may be applicable for the thin outer gap accelerator of the
Vela pulsar.  By ignoring the derivative of the potential field
along the magnetic field line, we reduce the two-dimensional Poisson
equation~(\ref{Poisson0_3}) to one-dimensional form as
\begin{equation}
\frac{\partial^2{}}{\partial{z^2}}\Phi'(x,z,\phi_p)=-4\pi[\rho(x, z, \phi_p)-\rho_{GJ}(x, \phi_p)].
\label{poisson1_3}
\end{equation}

At the lower boundary, where the accelerating electric field just begins, the boundary condition is
\begin{equation}
\Phi'(x, z=0, \phi_p)=0.
\end{equation}
At the upper boundary, which is located on the surface of fixed magnetic
field line, we impose
\begin{equation}
\Phi'(x, z=h_2, \phi_p)=0~~ \mathrm{and}~~E_{\perp}(x, z=h_2, \phi_p)=0.
\end{equation}
Using the above boundary conditions and imposing that
 $\Phi'$ and  $\partial\Phi'/\partial z$ are continuous at the height $h_1$,
 we obtain the solution for the Poisson equation~(\ref{poisson1_3}) as
\begin{equation}
\Phi'(x, z, \phi_p)=-2\pi\left\{
   \begin{array}{lcc}
            \{\rho_1(x, \phi_p)-\rho_{GJ}(x, \phi_p)\}z^2\\+C_1z,  \\ $for$~0\leq z\leq h_1\\
            \{\rho_2(x, \phi_p)-\rho_{GJ}(x, \phi_p)\}\\(z^2-h_2^2(x, \phi_p))\\+D_1(z-h_2(x, \phi_p)) ,   \\ $for$~h_1\leq z\leq h_2
   \end{array},\right.
\label{potential_3}
\end{equation}
where
\[
C_1=\frac{(\rho_1-\rho_{GJ})h_1(h_1-2h_2)-(\rho_2-\rho_{GJ})(h_1-h_2)^2}{h_2},
\]
\[
D_1=\frac{(\rho_{GJ}-\rho_2)h_2^2+(\rho_1-\rho_2)h_1^2}{h_2},
\]
Because the two boundary conditions are imposed on the upper boundary ($z=h_2$),
 the position of the upper boundary cannot be chosen arbitrarily.
On the other hand, the condition $E_{\perp}(x, z=h_2, \phi_p)=0$ gives
the relation between the charge densities
($\rho_1$, $\rho_2$) and the gap thickness ($h_1$, $h_2$) as
\begin{equation}
(\rho_2-\rho_{GJ})h_2^2+(\rho_1-\rho_2)h_1^2=0.
\label{condi_3}
\end{equation}

The present model  expects that the charge density should 
 proportional to the Goldreich-Julian charge density in the screening 
region (see Wang et al. 2010),  and therefore we express
 $\rho(x,z,\phi_p)-\rho_{GJ}(x,\phi_p)\sim g(z,\phi_p)\rho_{GJ}(x,\phi_p)$,
where
\begin{equation}
g(z, \phi_p)=\left\{
   \begin{array}{ccc}
            -g_1(\phi_p), &$if$& 0\leq{}z\leq{}h_1\\
            g_2(\phi_p), &$if$& h_1<z\leq{}h_2
   \end{array}.\right.
\label{cdensity}
\end{equation}
We assume that  $g_1>0$ and $g_2>0$ so that   $|\rho|<|\rho_{GJ}|$
in the main region and $|\rho|>|\rho_{GJ}|$ for the screening region,
respectively.
Substituting equation~(\ref{cdensity}) into equation~(\ref{condi_3}),
we obtain
\begin{equation}
(\frac{h_2}{h_1})^2=1+\frac{g_1}{g_2}
\label{eqn9}
\end{equation}

We approximate that  the Goldreich-Julian charge density
as $\rho_{GJ}(x)\approx -\frac{\Omega Bx}{2\pi cs}$ (Cheng, Ho Ruderman 1986a),
 where $\Omega$ is the angular frequency of the pulsar.
Using  the relations that  $\partial(Bh^2_2)/\partial{}x\sim{}0$, $\partial(z/h_2)/\partial{}x\sim{}0$, $\partial(h_1/h_2)/\partial{}x\sim{}0$, and $\partial{}s/\partial{}x\sim{}0$,
the accelerating electric field, $E_{||}=-\partial\Phi'/\partial{}x$ is written
down as
\begin{equation}
E_{||}(x, z, \phi_p)\sim\frac{\Omega B}{cs}\left\{
\begin{array}{lcc}
-g_1z^2+C'_1z, \\ $for$~0\le{}z\le{}h_1 \\
g_2(z^2-h^2_2)+D'_1(z-h_2), \\ $for$~h_1<z\le{}h_2
\end{array},
\right.
\label{electric_3}
\end{equation}
where
\[
C'_1=-\frac{g_1h_1(h_1-2h_2)+g_2(h_1-h_2)^2}{h_2},
\]
\[
D'_1=-\frac{g_2h_2^2+(g_1+g_2)h_1^2}{h_2}.
\]

In the three dimensional rotating magnetic field structure,
 the thickness of the gap $h_2$ is a function of the distance
to the star along the field line $x$.  We define the gap fraction $f$
 measured  on the stellar surface as (Wang et al., 2010),
\begin{equation}
f\equiv\frac{h_2(R_s)}{r_p},
\label{def_fm}
\end{equation}
where $R_s$ is the stellar radius, and $r_p(\phi_p)$ is the
polar cap radius and depends on polar angle in three-dimensional magnetic field geometry.
Note that because the electric field  $E_{||}$ is proportional to $Bh^2_2$,
which is almost constant along the field line for the dipole field,
it can be found that $E_{||}\propto{}f^2$. This indicates that
the strength of the electric field and resultant emissivity of the curvature
radiation increase with the fractional gap thickness.

The height($z$) measured from the
last-open field lines is a function of distance ($x$) along the magnetic
field line.  In order to make it easy to distinguish the two layers
in any $x$, we introduce the factor $a$ to represent the magnetic field lines at a given layer
and take $a=1$ for the last-open field lines. {{}We first determine the coordinate values  $[X_0 (\phi_p), Y_0(\phi_p), Z_0(\phi_p)]$ of the last-open
field lines at the stellar surface. Then an arbitrary layer of the magnetic field lines with a given ``a'', whose coordinate values $[X'_0 (\phi_p), Y'_0(\phi_p), Z'_0(\phi_p)]$, can be determined by using $X'_0(\phi_p)=aX_0(\phi_p)$, $Y'_0(\phi_p)=aY_0(\phi_p)$ and $Z'_0(\phi_p)=(R^2_s-a^2X'^2_0-a^2Y'^2_0)^{1/2}$.}
 The relation between the $z$ and the $a$ is approximated as
\begin{equation}
z(x,\phi_p)=\frac{1-a}{1-a_{min}}h_2(x,\phi_p),
\end{equation}
where $a_{min}$ corresponds to the upper boundary of the gap. In this paper,
the dimensionless thickness $a_{min}$ is assumed to be
the same for  different polar angle $\phi_p$ because each layer should have a similar polar cap shape on the stellar surface, and is chosen as the fitting
parameter of the light curve. The thickness $a_{min}$ determines
the width of the pulse peaks and the phase separation between two main peaks.

\subsection{Curvature radiation spectrum}
Our calculation method for the phase-average spectrum and
phase-resolved spectra  is based on Tang et al. (2008). In a volume
element $\Delta{V}$, there are $\Delta{N}=\Delta{V}n$ particles accelerated by
the electric field that is described by equation~(\ref{electric_3}). The accelerated particles release  the power gained from
the accelerating electric field, $eE_{\parallel}c$, through
  the curvature radiation process.  The total radiation power for each particle is $l_{cur}(\vec{r})=2e^2c\gamma^4_e(\vec{r})/3s^2(\vec{r})$, where
$\vec{r}=(x,~z,~\phi_p)$. The typical
 local Lorentz factor $\gamma_e$ of the primary particles can be obtained
by requiring $eE_{\parallel}c=l_{cur}$, which gives
\begin{equation}
\gamma_e(\vec{r})=\lbrack\frac{3}{2}
\frac{s^2(\vec{r})}{e}E_{\parallel}(\vec{r})\rbrack^{1/4}.
\end{equation}

The photon spectrum at each position of radiation  is
\begin{equation}
\frac{d^2 N_{\gamma}(E_{\gamma}, \vec{r})}{d E_{\gamma} d t}=\frac{\Delta{N(\vec{r})}
\sqrt{3}e^2\gamma_e}{2\pi{\hbar}sE_{\gamma}
}F(\chi),
\end{equation}
where $\chi=E_{\gamma}/E_{cur}(\vec{r})$, $E_{cur}(\vec{r})
=(3/2)\hbar{}c\gamma_e^3(\vec{r})/s(\vec{r})$ is the characteristic energy of the radiated curvature photons, $s(\vec{r})$ is curvature radius, and
$F(\chi)=\int^{\infty}_{\chi}K_{5/3}(\xi)d\xi$, where $K_{5/3}$ is the modified Bessel functions of order 5/3. {{}The local solid angle of curvature photon $\Delta{\Omega}\simeq\pi\theta^2$}, where
\begin{equation}
\theta(\vec{r})=Max(\Delta{l}/s(\vec{r}), \gamma(\vec{r})mc^2/eB(\vec{r})s(\vec{r})),
\label{omega}
\end{equation}
where the first term represents
 the solid angle caused by the curvature of the magnetic field line
and the second term is the minimum  emission angle measured
from the direction of the magnetic field line. Here $\Delta{l}$ is the gird size of the cell along field line.

With equation~(\ref{cdensity}), we express the number density at each point as
$n(\vec{r})=\frac{\Omega B}{2\pi c} [1-g(z,\phi_p)]$.
The volume element at each position  can be calculated
\begin{equation}
\begin{array}{ccc}
\Delta{V(\vec{r})}=\frac{2{\pi}r_{p}h(R_s)B(R_s)\Delta{l}(x)}{N_AN_BB(\vec{r})}
\end{array},
\label{volum_3}
\end{equation}
where $2{\pi}r_{p}h_2(R_s)B_s$
is the total magnetic flux through in the gap, and $N_A$ and $N_B$ are
the number of the grids in the direction of the trans-field direction (z-direction) and the azimuthal direction, respectively.

The total photon flux received at Earth is
\begin{equation}
F_{tot}(E_{\gamma})=\frac{1}{D^2}\sum_{\vec{r}_i}\frac{1}{\Delta{\Omega}(\vec{r}_i)}\frac{d^2 N_{\gamma}(E_{\gamma}, \vec{r}_i)}{d E_{\gamma} d t}
\label{spectrum}
\end{equation}
where $\vec{r}_i$ represents the position of the i$^{th}$ cell, from which the emission
can be observed, and  $D$ is the distance to the pulsar from Earth. We note
 that because $r_i$ depends on the viewing angle and the
inclination angle, the spectrum~(\ref{spectrum}) is a function of the
viewing angle and the inclination angle.

To calculate the phase-resolved spectra and the light curves,
the arrival times of the photons are binned by pulse phase. For example,
Figure~\ref{skymap} represents photon-mapping on the plane spanned by
the viewing angle and the pulse phase
 using the rotating dipole fields (section~\ref{geometry}).
 For each viewing angle, the number of the photons
 measured at pulse phases between $\psi_1$ and $\psi_2$ is calculated from
\begin{equation}
N_{\gamma}(E_1, E_2, \psi_1, \psi_2)\propto\int^{E_2}_{E_1}{F_{tot}(E, \psi_1 \leq \psi \leq \psi_2)}dE.
\label{int_spec}
\end{equation}

\subsection{Emission Geometry}
\label{geometry}
It has been considered that the positions of the first and the second peaks of
 the pulsar are determined by the geometry of the magnetic field. In this study,  we adopt the  rotating vacuum dipole field  to calculate the light curves and spectra. For the rotating dipole, the magnetic field $\vec{B}(r)$ is given by
\begin{equation}
\vec{B}=\hat{r}[\hat{r}\cdot(\frac{3\mu}{r^3}+\frac{3\dot{\mu}}{cr^2}+\frac{\ddot{\mu}}{c^2r})]-(\frac{\mu}{r^3}+\frac{\dot{\mu}}{cr^2}+\frac{\ddot{\mu}}{c^2r})
\end{equation}
(Cheng, Ruderman \& Zhang, 2000), where $\dot{\mu}=\mu(\hat{x}\sin{\alpha}\cos{{\Omega}t}
+\hat{y}\sin{\alpha}\sin{{\Omega}t}+\hat{z}\cos{\alpha})$ is the magnetic moment vector, $\hat{r}$ is the radial unit vector, and $\alpha$ is the inclination angle. The Runge-Kutta method is employed to trace out  field lines and
to find the polar cap rim  $[X_0(\phi_p), Y_0(\phi_p), Z_0(\phi_p)]$.

Once the upper boundary  $a_{min}$ is chosen, we trace the open field lines defining the outer gap, and calculate the emission direction
in the observer's frame  at each point on the field lines.
In Tang et al. (2008), the emission direction coincides with the direction of
 the rotating vacuum dipole field defined in the co-rotating frame.
 In the present  three-dimensional  model, on the other hand, the emission direction calculated in only observer's frame with the computation method employed
by  Takata et al. (2007). The curvature photons are assumed to be emitted in the direction of the particle motion, which can be described as
\begin{equation}
             \vec{v}=v_p\vec{B}/B + \vec{r}\times\vec{\Omega}
\end{equation}
where the first term represents the motion along the magnetic field line, $v_p$ is calculated from the condition that $|\vec{v}|=c$, and the second term is the drift motion. The polar angle to the rotation axis $\zeta$ of the emission direction
 and the pulse phase  $\psi$  are calculated from
\begin{equation}
\left\{
   \begin{array}{ccc}
             \cos{\zeta}=v_z/v\\
             \psi=-\cos^{-1}(v_x/v_{xy})-\vec{r}\cdot\vec{\hat{v}}/R_L
   \end{array}\right.
\end{equation}
(Yadigaroglu 1997). If a point in a field line satisfies $|\zeta-\beta|<\varphi(r)$, the radiation of that point can be seen by the observer, where
 $\beta$ is the viewing angle and $\varphi(\vec{r})=\Delta{l}(\vec{r})/s(\vec{r})$.
Because  the positions of the first  and second peaks  in the light curve
is also determined by viewing geometry, the predicted inclination angle $\alpha$ and the viewing angle $\beta$ are uniquely  determined by
 comparing the calculated  peak separation with the observations.

\section{Results and Discussion}
In this section, we apply our model to the Vela pulsar.
In our  two-layer model, the general  properties of  $\gamma$-ray
emissions from the outer gap    are characterized
 by $f$, $g_1$ and $h_1/h_2$.
Specifically, the gap fraction $f$ mainly determines
the cut-off energy in the spectrum.
 Zhang \& Cheng (1997) suggested a self-consistent mechanism provided
 by the X-rays emitted by the polar cap that is heated by the return current
to restrict the gap. Their fractional size of the gap,
$f_{ZC}=0.32P^{26/21}_{-1}B^{-4/7}_{12}$, predicts that of the Vela
pulsar is 0.15.  In the two-dimensional model (Wang et al. 2010),
 $f=0.16$ was used for fitting the phase-averaged spectrum of the Vela pulsar.
The current in the main acceleration region, $1-g_1$, and the ratio
 between the thicknesses of the main acceleration region and that of
 the whole gap, $h_1/h_2$, determine the photon index of the spectrum together. For the Vela pulsar,
the fitting of the two-dimensional two-layer model gives $1-g_1=0.08$ and $h_1/h_2=0.927$ (Wang et al. 2010).

\begin{figure}
\centerline{\psfig{figure=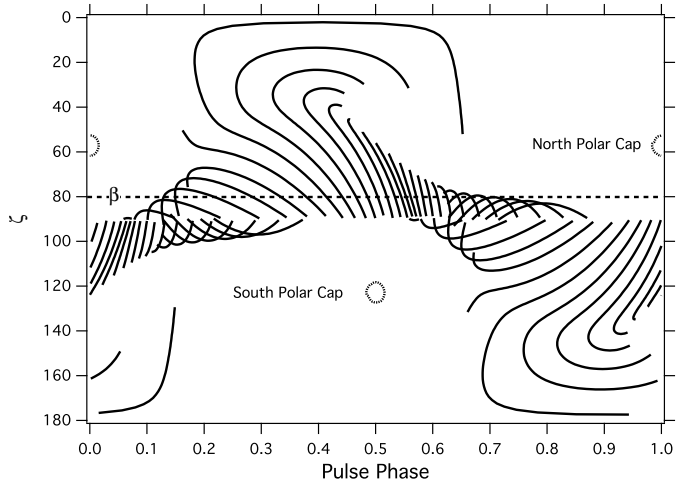, width=9.2cm, clip=}}
\caption{The skymap of the radiations outside of the null charge surface with inclination angle $\alpha=57^{\circ}$. The $x$-axis is the pulse phase and the $y$-axis is the direction of the radiation. {{}The dashed line is the viewing angle, which is chosen as $\beta=80^{\circ}$}.}
\label{skymap}
\end{figure}

The position of the upper boundary
 $a_{min}$ is treated as a model parameter, because it is affected  by
the actual magnetic field structure, which is not understood well.
 Together with the inclination angle $\alpha$
and the viewing angle $\beta$,  the  position of the upper boundary
 $a_{min}$ can be constrained by the detailed structure of
pulsed profile.
As we show in the following section, we find  that the set of
($\alpha,~\beta,~a_{min})=(57^{\circ},~80^{\circ},~0.935$) reproduces
well both the light curves and the spectra.  Figure~\ref{skymap} shows
the sky-map of the emitted photons from the magnetic surface of $a=0.95$ with
$\alpha=57^{\circ}$ and $\beta=80^{\circ}$. Although there are many sets of the inclination angle and the viewing angle that can provide the observed peak separation, our choice of these two angles is not arbitrary. The reason for this will be discussed later.

\subsection{A Simple Case}
Firstly, we assume
 constant $1-g_1$, $h_1/h_2$ and $f$ in the azimuthal direction.
By fitting the phase-averaged spectrum for the Vela pulsar,
 we obtain $1-g_1=0.05$, $h_1/h_2=0.927$ and $f=0.2$ for the present
three-dimensional model. Here we assume the distance of  $D=325$~pc.

\begin{figure}
\centerline{\psfig{figure=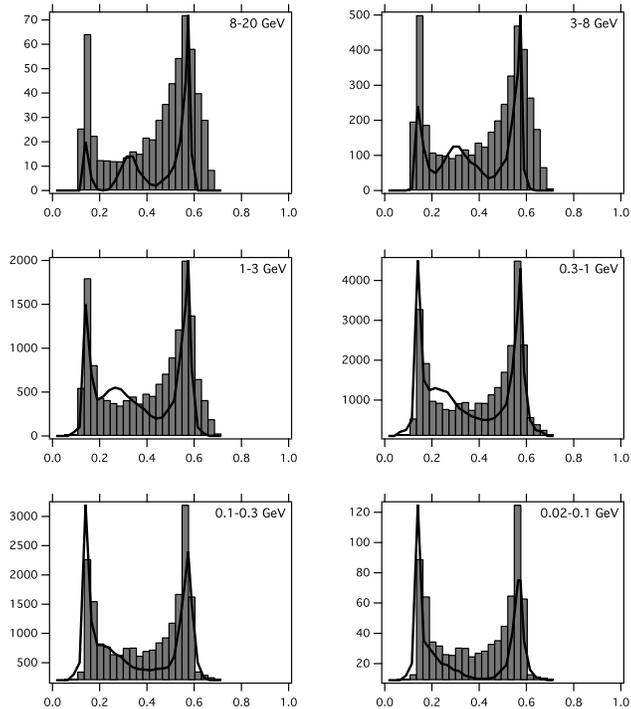, width=9.2cm, clip=}}
\caption{The energy dependent light curves (histogram) with $1-g_1=0.05$, $h_1/h_2=0.927$ and $f=0.2$. The solid lines are the observed light curves from Fermi-LAT (Abdo et al., 2010b).{{}The Y-axis is the counts of the observed light curve. }}
\label{edlc_nd}
\end{figure}

Figure~\ref{edlc_nd} compares the calculated (histograms) and the observes
(solid lines) pulse profiles  in six different energy bands. We can
see in Figure~\ref{edlc_nd} that the model provided pulse profiles are generally
consistent with the observations.  For example, the pulsed profiles in
wide energy bands have two main peaks with the  separation of  about 0.42.
The positions of the two peaks in the light curves  are determined by
 the geometry of the magnetic field.

In the present simple model, however,
the calculated pulse profile can not explain the existence of the observed
third peak between two main peaks,  as we can see in Figure~\ref{edlc_nd}.
The  peaks emerging in the calculated  pulse  profiles of Figure~\ref{edlc_nd}
are caused by so called  caustic effects (Yadigaloglu and Romani 1995;
 Cheng, Ruderman and Zhang 2000; Dyks, Harding and Rudak  2004), in which
 more photons are observed at  narrow width of the rotation phase  due to
the special-relativistic effect (that is, the
aberration of the emission direction and photon's travel time).
In the caustic model, the phases of the peaks  are determined
by the magnetic field structure and they do not depend on the energy band.
On the other hand,  the observed phase of the  third peak shifts
 with the energy bands.
The simple caustic model can not explain the
phase shift of the observed third peak.
On these ground,
 we speculate that the existence of the third peak and its phase
shift is related with  more complex structure of the emission region.

\subsection{More Realistic Case}
In this section,
we discuss  the third peak with a  more complex  gap structure
by taking into account   the azimuthal structure
of  the fractional gap thickness $f$, the ratio of the thicknesses of
the main acceleration region to the whole gap thickness, $h_1/h_2$, and
the particle number density in the main acceleration region, $1-g_1$.

\subsubsection{Fractional gap thickness}
Firstly, we consider the  azimuthal structure of fractional gap thickness
and compare the calculated light curves with the observations.

By the  definition of the gap fraction of  equation~(\ref{def_fm}),
 $f\propto\frac{1}{r_p}$,  we may
choose the form of the azimuthal distribution of $f$ as,
\begin{equation}
f(\phi_p)=\frac{C}{r_p(\phi_p)},
\label{def_f}
\end{equation}
where the $C=0.18r_{p}^{max}$, $r_{p}^{max}$ is the maximum value of the
 polar cap radius, and the factor 0.18 is chosen
by fitting the phase averaged-spectrum.

\begin{figure}
\centerline{\psfig{figure=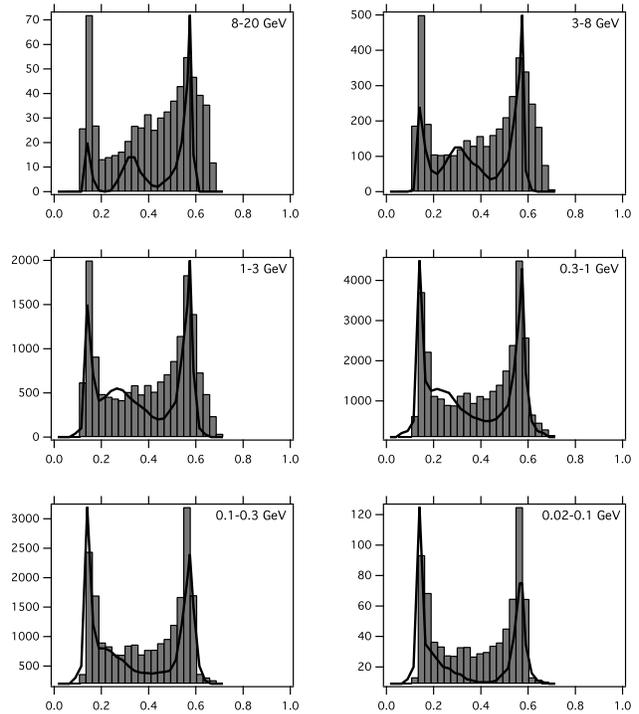, width=9.2cm, clip=}}
\caption{The energy dependent light curves (histogram) with the distribution of $f$ that provided by the three dimensional outer gap. $1-g_1$ and $h_1/h_2$ are chosen as 0.05 and 0.927 respectively. The solid lines are the observed light curves from Fermi-LAT (Abdo et al., 2010b).}
\label{edlc_f}
\end{figure}

The pulse profiles using the fractional thickness  $f$
described by equation (\ref{def_f}) are shown in Figure~\ref{edlc_f},
where  $g_1=0.05$ and $h_1/h_2=0.927$ are the same with those
of Figure~\ref{edlc_nd}. By comparing Figure \ref{edlc_nd} and \ref{edlc_f}, we can find the effects of azimuthal
distribution of $f$ on the pulse profiled. For example, the azimuthal
structure of the fractional gap thickness produces more bridge emissions
as well as a third-peak-like feature at the phase $\sim 0.3$
 of the pulse profile (in particular for higher energy bands) 
 in Figure~\ref{edlc_f}.

\begin{figure}
\centerline{\psfig{figure=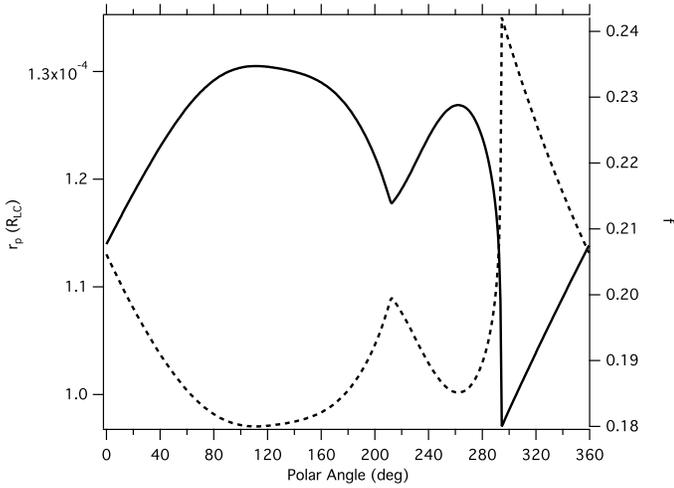, width=9.2cm, clip=}}
\caption{The distributions of the polar cap radius $r_p$ (solid line) and the resultant $f$ (dashed line)
as a function of  Polar Angles, with the vacuum dipole field. {{}The definitions of the polar angle=0$^{\circ}$ and 180$^{\circ}$ have been made in the text.}}
\label{rp_phi}
\end{figure}

Figure~\ref{rp_phi} shows   the polar cap radius $r_p$ (solid-line) and
the resultant fractional gap thickness $f$ (dashed-line) as a function of
the polar angle. 
{{}The polar angle=0$^{\circ}$ is defined at the right-hand side of the north pole's magnetic axis in the plane including the rotation axis and the magnetic axis, where the rotation axis is at the left-hand side of magnetic axis in the north pole. Hence the polar angle=180$^{\circ}$ is defined at the left-hand side of the magnetic axis. The angle between the magnetic axis and the rotation axis is chosen to be $<$ 90$^{\circ}$, therefore the outgoing radiation from the south pole can be detected with viewing angle $<$ 90$^{\circ}$.}
In the south pole, the emissions
on the magnetic field lines emerging from polar angle larger  (or smaller)
than $180^{\circ}$ produces the first (or second) peak.
 As shown in Figure~\ref{rp_phi}, the fractional thickness
$f$ becomes maximum around the polar angle of 200$^{\circ}$,
where $r_p(\phi_p)$ becomes minimum.
Because larger fraction thickness $f$ produces a stronger $E_{||}$
as equation (10) indicates, the  emissions on the magnetic field lines
 emerging from $\sim$ 200$^{\circ}$ is more stronger than
those from $\sim 160^{\circ}$ and from $\sim 240^{\circ}$. Consequently,
  third-peak-like  structure is formed in the light curves of
Figure~\ref{edlc_f}.

\subsubsection{Gap thickness of the two layers}
\label{gapthick}
In the electrodynamic  point of view, the thickness of the screening region
of the gap, $h_2-h_1$,   will be  determined by the photo-photon
 pair-creation rate between the $\gamma$-rays and X-rays from the
 stellar surface. It is expected that
as the null charge surface is closer
 to the stellar surface, the  number density of the X-ray photons increases
in the gap and therefore the screening region becomes thinner.
This indicates that the ratio of the thicknesses of the main acceleration
region to the whole gap thickness,  $h_1/h_2$,  is closer to unity as the
 null charge surface approaches to the stellar surface. Because
the radial distance to the null charge surface is a function of the
azimuthal coordinate, it is expected that the ratio $h_1/h_2$ varies in the
azimuthal direction.  To simulate above dependency,
let's assume the formula of the ratio as
\begin{equation}
\frac{h_1}{h_2}(\phi_p)=B_1+B_2\frac{1/r_{null}(\phi_p)-1/r_{null}^{max}}{1/r_{null}^{min}-1/r_{null}^{max}},
\label{h1/h2}
\end{equation}
where the factor $B_1$ represents the ratio at the
polar angle, at which the radial distance to the null charge surface
becomes  maximum, $r_{null}^{max}$, and $B_1+B_2$ corresponds to
 the ratio at azimuthal  angle, at which the radial distance
 to the null charge surface becomes  minimum, $r_{null}^{min}$. By fitting
the spectral shape, we obtain $B_1=0.89$ and $B_2=0.09$.

\begin{figure}
\centerline{\psfig{figure=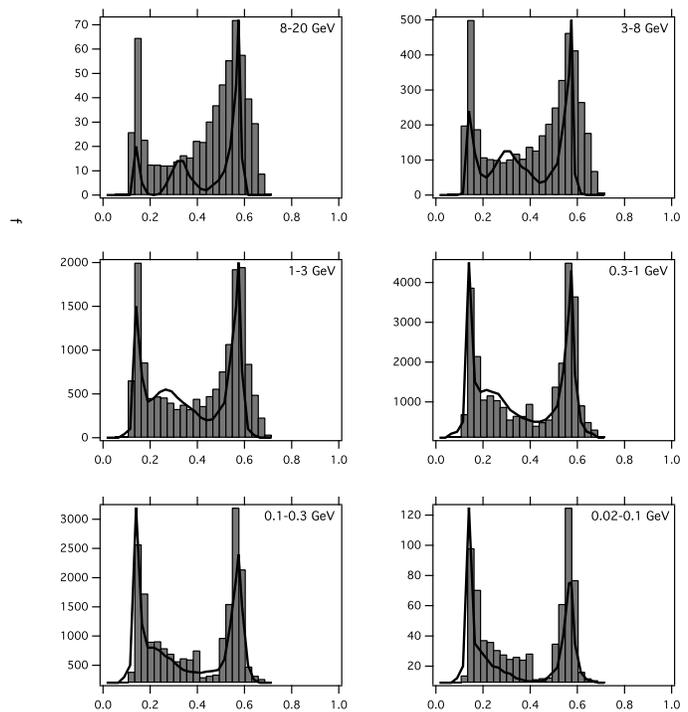, width=9.2cm, clip=}}
\caption{The energy dependent light curves (histogram) with the distributions of $h_1/h_2$ that provided by the three dimensional outer gap. $1-g_1$ and $f$ are chosen as 0.05 and 0.2 respectively. The solid lines are the observed light curves from Fermi-LAT (Abdo et al., 2010b).}
\label{edlc_h}
\end{figure}

To see the effects of the azimuthal dependency of the thickness ratio,
$h_1/h_2$, on the pulse profiles,  Figure~\ref{edlc_h} shows
the simulated pulse profiles with equation~(\ref{h1/h2}) and the fixed
fractional thickness  of $f=0.2$ and the number density of  $1-g_1=0.05$,
which are the same with those of Figure~\ref{edlc_nd}.
By comparing Figures~\ref{edlc_nd} and~\ref{edlc_h}, one can see that
 the distribution of the $h_1/h_2$ does not affect much
 the light curves above energy  $E>1$GeV, but it makes a bump in
the bridge ($\sim 0.2-0.3$ pulse phase) of the light curve around 100MeV.
 We can see those light curves
below $E<1$~GeV in Figure~\ref{edlc_h} are  more consistent
  with the observation than those of Figure~\ref{edlc_nd}.
Wang et al. (2010) show that the ratio of the thicknesses $h_1/h_2$
affects the  emissivity in the screening region. Because the particles in
 the screening regions, where the accelerating electric field is small,
 mainly produces the curvature photons of the energy less than
 $E\le 1$~GeV,  the azimuthal distributions of the thickness ratio
affects the light curves below $E\le 1$~GeV.

\begin{figure}
\centerline{\psfig{figure=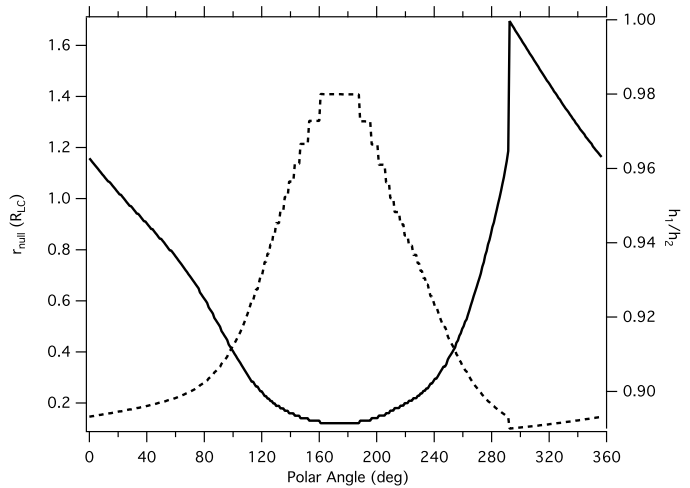, width=9.2cm, clip=}}
\caption{The distributions of the distance of the null charge surface from stellar surface (solid line) and the resultant $h_1/h_2$ (dashed line). }
\label{h1h2_phi}
\end{figure}

Figure~\ref{h1h2_phi} shows the distributions of the radial  distance to
 the null charge surface $r_{null}$ (solid-line) for
the inclination angle $\alpha=57^{\circ}$ and
the corresponding $h_1/h_2$ (dotted-line) expressed by equation~(\ref{h1/h2}).
It shows that, at the polar angle$\sim$170$^{\circ}$ that corresponds to pulse
phase$\sim$0.4, the screening region is {{}thinnest, which means the radiations
below 1GeV at the pulse phase $\sim$ 0.4 are weaker than} those of other
 pulse phases. Consequently, the shape of the bridge emissions of
the light curves  becomes  more consistent  with
the observations and stands out the third-peak-like structure.

\subsubsection{Number density}
The azimuthal distribution of the particles will be  more complicated,
because it is related with (1) the motion of the particles under
 the Lorentz force
 and (2) the propagating direction of the curvature photons, which are
eventually converted into the pairs in the gap.  For example, the
non-co-rotational electric field perpendicular to the magnetic field lines
causes $\vec{E}\times\vec{B}$ drift motion of the particles. Moreover,
because of the flight time of the photons and of
  the rotation of the magnetosphere, the magnetic field line, on  which
the pairs are created by the pair-creation process,  is different from the
 field line, where the photons were emitted by the curvature radiation process.

With the two-dimensional two-layer model,
 Wang et al. (2010) found
  that the phase-averaged spectra for most of the $\gamma$-ray pulsars
 can be reproduced  by the averaged charge density of
$\bar{\rho}_0\equiv [h_1\rho_1+(h_2-h_1)\rho_2]/h_2\sim 0.5$. We want to point
 out that because  the actual $E_{\perp}(\phi_p)$ is different
 at different $\phi_p$, particles in different $\phi_p$-cell 
may drift into other $\phi_p$-cell. Therefore we assume
that the averaged charge density has a distribution in
 the azimuthal direction due to the effect of $E_{\perp}\times B$
 drift motion.
 Without the effect of drifting, the number of particle in $\phi_p$-cell is $N_0(\phi_p)\propto{}f(\phi_p)\bar{\rho_0}(\phi_p)$. With the drifting effect, the real number of particle in $\phi_p$-cell ($N(\phi_p)$) may come from $(\phi_p+\Delta{\phi_p})$-cell, i.e. $N(\phi_p)=N_0(\phi_p+\Delta{\phi_p})$. Since the average charge density $\bar{\rho}(\phi_p)=N(\phi_p)/f(\phi_p)$, we obtain
\begin{equation}
\bar{\rho}(\phi_p)=\bar{\rho}_0\frac{f(\phi_p+\Delta{\phi_p})}{f(\phi_p)},
\end{equation}
where $\bar{\rho}_0=0.5$ is the averaged charge density without
drift motion. It may be expected that displacement of the particles due to
the drift motion becomes more important on the magnetic field line that has
a smaller radial distance to the  null charge surface,
 because the particles run
longer distance in the outer gap, which is extending the null charge surface
 to the light cylinder. To take into account such effect, we
may assume the formula of  displacement as
\begin{equation}
\Delta{\phi_p}=F\frac{1/r_{null}(\phi_p)-1/r_{null}^{max}}{1/r_{null}^{min}-1/r_{null}^{max}},
\end{equation}
where $F$ is a fitting parameter, which is chosen as -28$^{\circ}$.  We take
$\Delta{\phi_p}=0$ for $r_{null}^{max}$, because the null charge surface is
close to the light cylinder, therefore particles from other cells cannot drift in.
The negative value (that is, counter-direction of the rotation)
of the displacement may be expected because (1) more particles
are concentrated  around the upper part of the gap and (2) the direction of the
non-corotational electric field $E_{\perp}$ is negative value at upper part
 of the gap, indicating the direction of the drift motion is counter
direction of the rotation.

\begin{figure}
\centerline{\psfig{figure=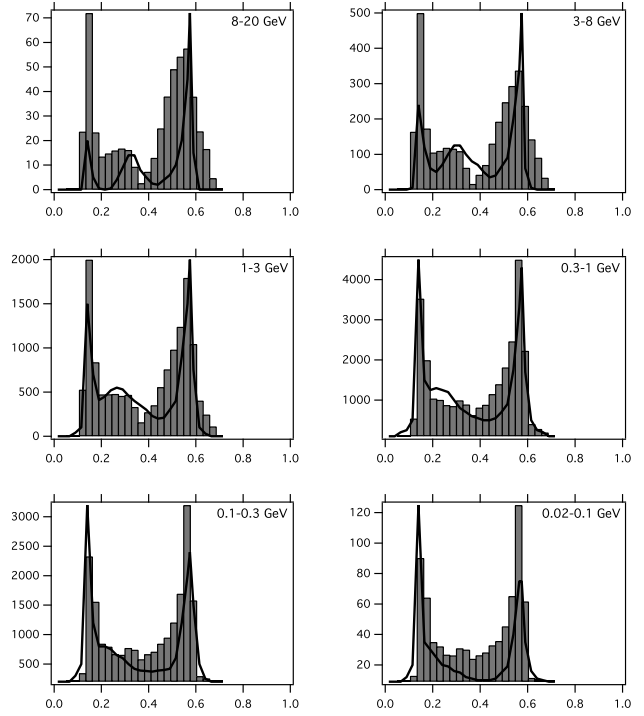, width=9.2cm, clip=}}
\caption{The energy dependent light curves (histogram) with the distributions of $1-g_1$ that provided by the three dimensional outer gap. $h_1/h_2$ and $f$ are chosen as 0.927 and 0.2 respectively. The solid lines are the observed light curves from Fermi-LAT (Abdo et al., 2010b).}
\label{edlc_rho}
\end{figure}

\begin{figure}
\centerline{\psfig{figure=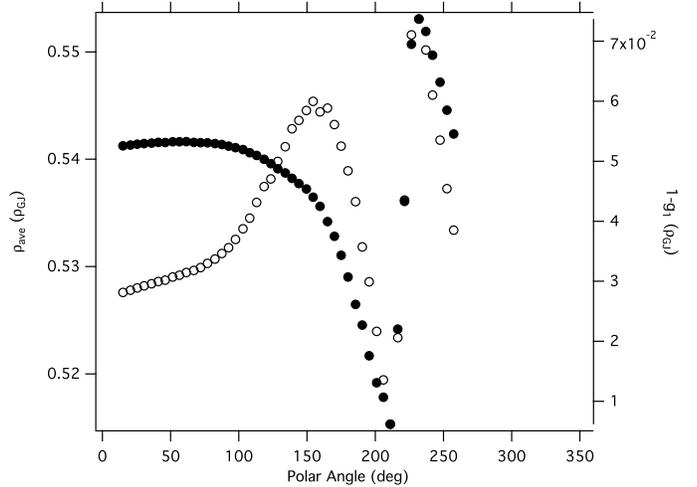, width=9.2cm, clip=}}
\caption{The distributions of the average charge density of the two layers (filled-circle) and the resultant $1-g_1$ (circle) of the field fields of different Polar Angles. }
\label{rho_phi}
\end{figure}

Using  the relationship between $g_1$, $g_2$ and $h_1/h_2$ given by equation~(\ref{eqn9}) and $\bar{\rho}= [h_1\rho_1+(h_2-h_1)\rho_2]/h_2$,
 the distribution of $1-g_1$ can be obtained.  Figure~\ref{edlc_rho}
is the energy dependent light curves calculated using the azimuthal
 distribution of $1-g_1$, where $h_1/h_2$=0.927 and $f$=0.2 are
 the same with those of Figure \ref{edlc_nd}.
Compared with Figure \ref{edlc_nd}, it shows that the distribution
of $1-g_1$ makes a third peak at the bridge region ($\sim 0.3$ pulse phase)
when the energy$>1$GeV. This is due to the distribution of the number density 
in the main acceleration region,
where the photons of $E>$1 GeV are emitted.
Figure~\ref{rho_phi} shows  the average charge density (filled-circle)
 and the resultant charge density in the main acceleration region (circle).
The average charge density and therefore the number density in the main 
acceleration region  acquire
the  maximum vale at polar angle around 240$^{\circ}$, which
corresponds to the pulse phase around 0.3.

\begin{figure}
\centerline{\psfig{figure=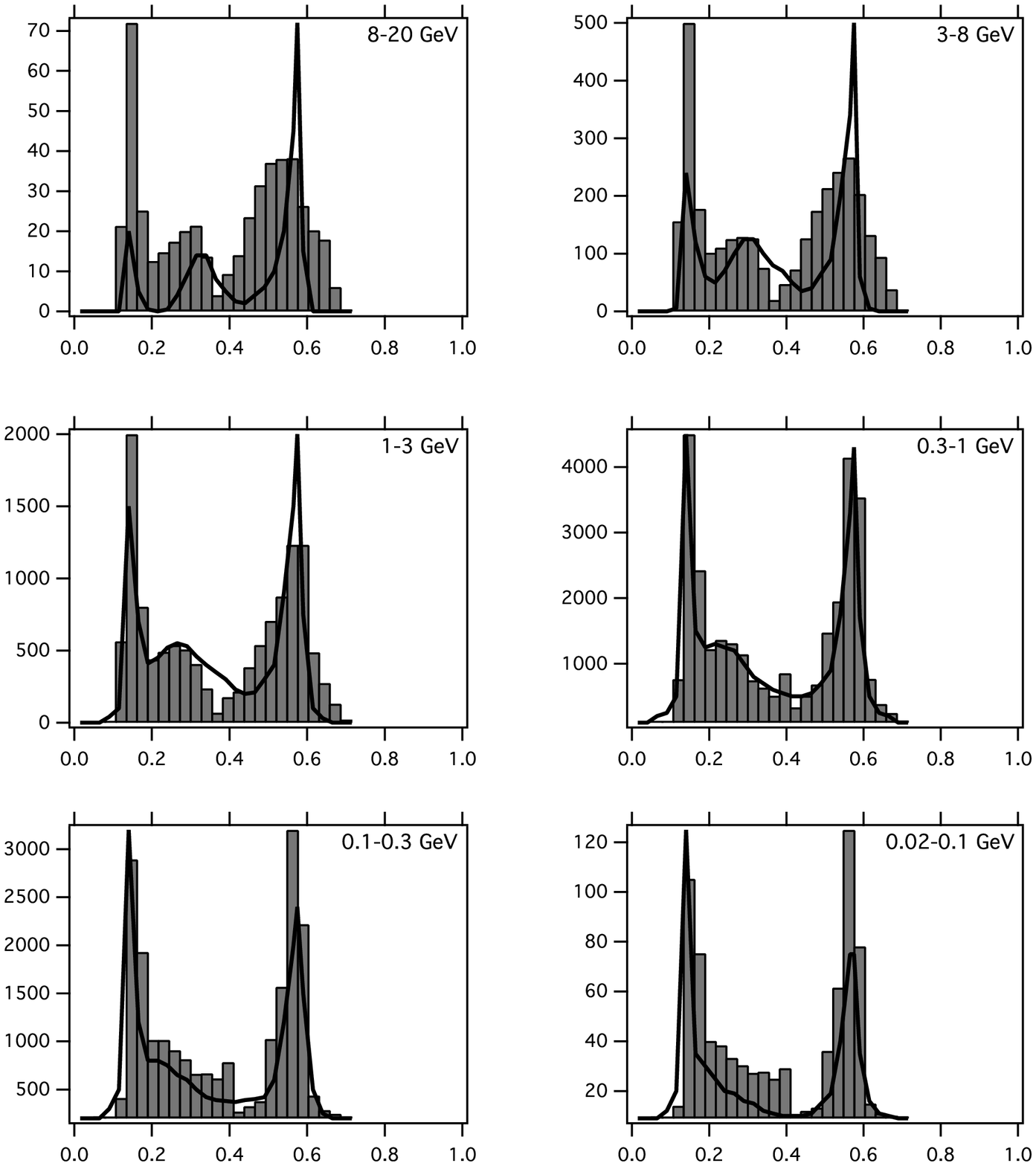, width=9.2cm, clip=}}
\caption{The energy dependent light curves (histogram) with the distributions of $1-g_1$, $h_1/h_2$ and $f$, provided by the three dimensional outer gap. The solid lines are the observed light curves from Fermi-LAT (Abdo et al., 2010b).}
\label{edlc_3dis}
\end{figure}
\begin{figure}
\centerline{\psfig{figure=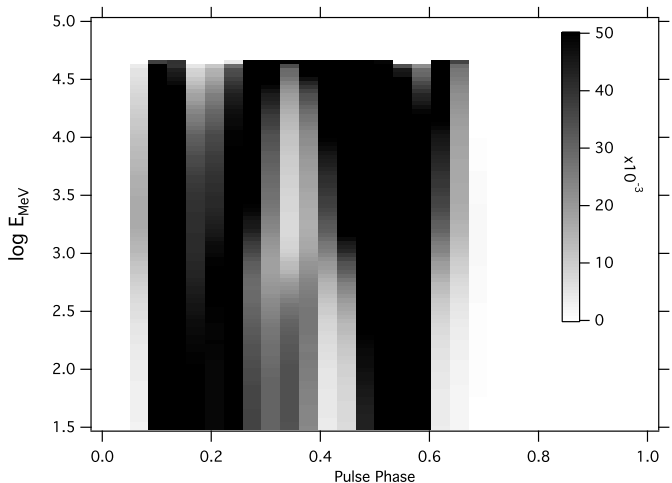, width=9.2cm, clip=}}
\caption{Intensity  map in  the pulse phase and  energy plane. The darkness represents the percentage of the number of the photons of certain interval of pulse phase in the total number of photons of certain interval of energy.}
\label{colmap}
\end{figure}

\subsection{Combined pulsed profile}
We have discussed the effects of the distributions of $f$, $h_1/h_2$ and
$1-g_1$ on the shape of the energy dependent light curves.
 As discussed in section~\ref{gapthick},  the distribution of thickness
 ratio $h_1/h_2$  mainly affects the light curve of $E\le 1$~GeV energy bands,
and  it makes a bump at pulse phase around 0.25 in the light curve below 1~GeV.
 The shape of bridge emission in the light curves above 1~GeV are mainly
affected  by the azimuthal distributions of the fractional thickness
$f$ and of the number density $1-g_1$. We demonstrated the third-peak-like
structure can be  produced by  the azimuthal structure of $f$, $h_1/h_2$ and
$1-g_1$.

Figure~\ref{edlc_3dis} shows the pulsed profiles calculated
by taking into account the azimuthal distributions for all
 $f$, $h_1/h_2$ and $1-g_1$.  We can see in Figure~\ref{edlc_3dis} that there is
a bump at the trailing part of the first peak in the pulse profile below 1~GeV,
 and it  shifts towards
the second peak in higher energy bands. Figure~\ref{colmap} {{}is} the intensity map in the pulse phase and the energy plane. The darkness
represents the number of the photons at the certain interval of the pulse phase. {{}It shows
that the third peak in the bridge region shifts towards the second peak as increases of the photon energy}.

The reason for the shift of the third peak is explained as follows.
In the light curves in the energy bands lower than 1~GeV, the distributions
of the thickness ratio $h_1/h_2$ makes a third-peak structure at $\sim0.2$
pulse phase. In the energy bands higher than 1~GeV, on the other hand,
 the azimuthal distribution of the fractional thickness, $f$, and the number
density, $1-g_1$, is more important, and they produce the third  peak
at $\sim 0.3-0.35$ pulse phase.
Consequently, the differences in the standing phases  of the
third peak due to the distributions  of $h_1/h_2$ and $f$ (or $1-g_1$)
 produce the shift of the third peak with the photon energy.

 Here the distribution of the polar radius $r_p$, position of null charge surface $r_{null}$, which determine the distribution of $1-g_1$, $h_1/h_2$ and $f$, depend on the inclination angle $\alpha$. For instance, if the inclination angle is small, such as 10$^{\circ}$, the shape of the polar cap is close to a circle, then the $r_p$ of different polar angles are nearly a same value, the third peak made by it is not obvious.  On the other hand, if the inclination angle is getting larger, the third peak made by the distribution of $f$ is getting more obvious. In order to provide appropriate distributions of the three parameters, we use inclination angle $\alpha=57^{\circ}$ and viewing angle $\beta=80^{\circ}$.

$Fermi$ $\gamma$-ray telescope have measured the $\gamma$-ray emissions from
about 60 pulsars. However, the Vela pulsar is the only one showing
 the shifting  third peak with energy. This may be because the Vela pulsar
is relatively young and its gap thickness is smaller than
other mature pulsars. For mature pulsars (e.g. the Geminga), the
gap thickness  will be much thicker ($f\sim 0.5$)
than that ($f\sim 0.2$) of the Vela pulsar. In such a thick gap,
the emissions from the different regions in the gap
 may smear  out the effects of azimuthal structure of the gap.
For the Crab pulsar,  the gap thickness $f\sim 0.1$
is smaller than that of  the Vela pulsar. However, the contributions
of the inverse-Compton scattering  from  the secondary pairs,
which are produced outside of the gap,
dominates the emissions in $\gamma$-ray energy bands (Tang et al. 2008).
This may be the reason that no third peak is observed in the  light
curves of the Crab pulsar.

\subsection{Spectra}
\begin{figure}
\centerline{\psfig{figure=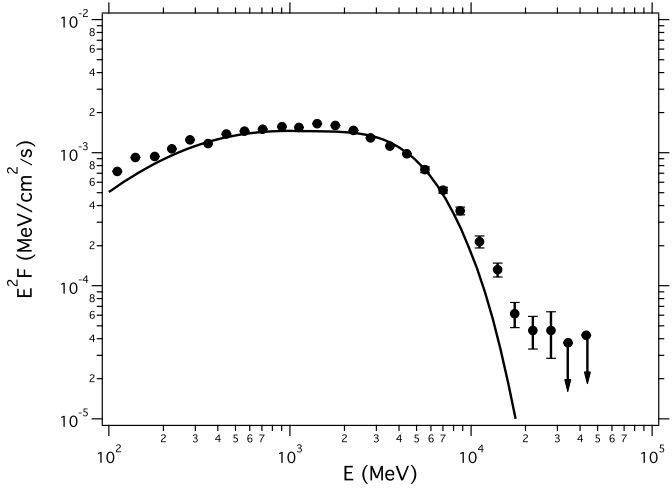, width=9.2cm, clip=}}
\caption{The phase averaged spectrum with the distributions of $1-g_1$, $h_1/h_2$ and $f$, comparing with the observed data (circle) from Fermi-LAT (Abdo et al., 2010b)}
\label{spec_3dis}
\end{figure}

\begin{figure}
\centerline{\psfig{figure=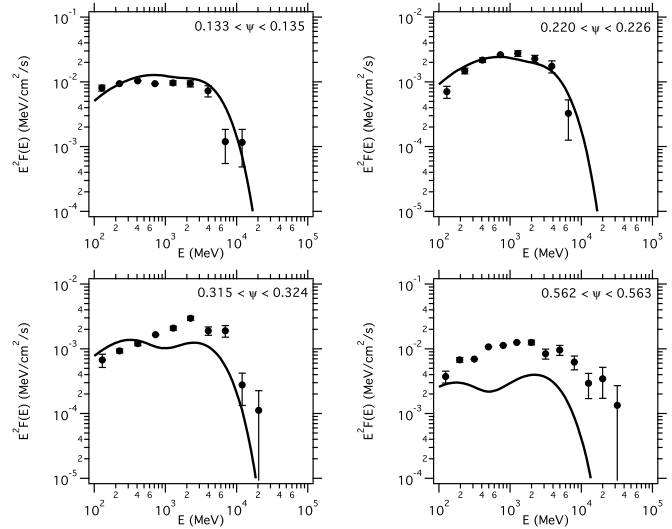, width=9.2cm, clip=}}
\caption{Four phase-resolved spectra provided by the more realistic model that using the distributions of $1-g_1$, $h_1/h_2$ and $f$, comparing with the observed data (circle) (Abdo et al., 2010b).}
\label{spec_pr}
\end{figure}

\begin{figure}
\centerline{\psfig{figure=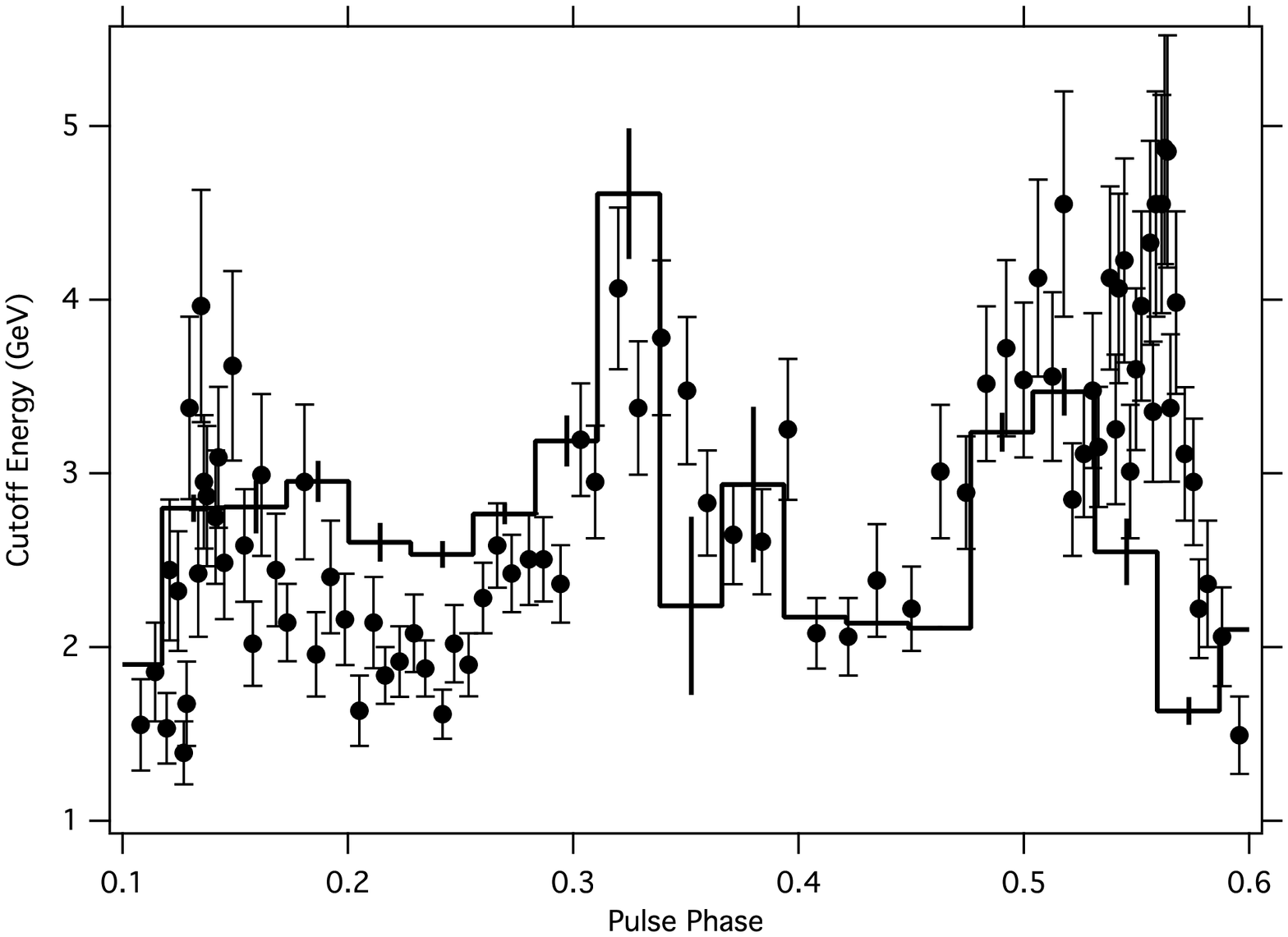, width=9.2cm, clip=}}
\caption{The cutoff energies of the phase-resolved spectra of different pulse phases given by the more realistic model, comparing with the observed data (circle) (Abdo et al., 2010b).}
\label{Ecut_Phi}
\end{figure}

\begin{figure}
\centerline{\psfig{figure=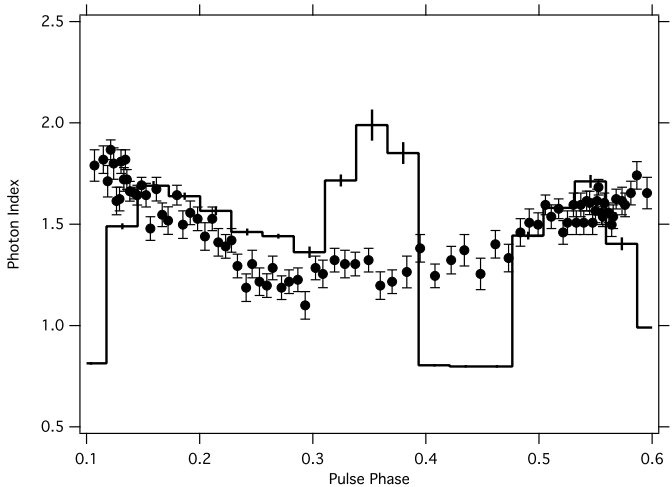, width=9.2cm, clip=}}
\caption{The photon indices of the phase-resolved spectra of different pulse phases given by the more realistic model, comparing with the observed data (circle) (Abdo et al., 2010b).}
\label{Gamma_Phi}
\end{figure}

The  spectral characteristics  are summarized in
Figures~\ref{spec_3dis}-\ref{Gamma_Phi}.  Figures~\ref{spec_3dis}
and~\ref{spec_pr}  compare the calculated (solid lines)  and
observed (filled-circles) phase-averaged
and phase-resolved spectra, respectively.
By fitting calculated spectra with power law plus exponential cut-off form,
 Figures~\ref{Ecut_Phi} and~\ref{Gamma_Phi} show
the cut-off energy and the photon index, respectively, as a function of the
pulse phase.

 As we can see from Figure~\ref{Ecut_Phi}, the present model can
explain the observed cut-off energy as a function of the pulse phase.
This can be explained as the effect of the distribution of the $f$, which decides the energies of the photons in the gap by determining the value of the accelerating electric field. And the cut-off energy of the spectrum is mainly determined by  the emission from the main acceleration region. From Figure~\ref{rp_phi}, in the observable region from around 0$^{\circ}$ to 260$^{\circ}$, there are three peaks of $f$, which correspond to the three peaks in Figure~\ref{Ecut_Phi}. And the pulse phase of the middle peak in Figure~\ref{Ecut_Phi} is consistent with that of the third peak in the light curve made by the distribution of $f$. These two facts indicate that the distribution of $f$ makes the calculated 
 cut-off energies of the phase-resolved spectra consistent 
with the observed one.

Because the photon index is much more sensitive to the shape of the spectrum than the cut-off energy, the Figure~\ref{Gamma_Phi} is not so consistent with the observation, where the photon indices between pulse phase of 0.3 to 0.5 are different with the observed ones. As the spectrum of $0.315<\psi<0.324$ in Figure~\ref{spec_pr} shows, the present model predicts a larger photon index at the 
pulse phase of 0.3-0.4 than 
the observations. For the pulse phase=0.4 to 0.5, as we can see from the energy dependent light curves (Figure~\ref{edlc_3dis}), the calculated photons are 
more than those of the observation. Consequently, the calculated curvature radiation from the main acceleration region is too strong, leading the calculated spectrum is hard with a photon index $\sim 1$.   For the other pulse phases, as shown by the Figure~\ref{spec_pr}, the flux and the shape of the spectra are close to the observed ones, and the photon indices of these pulse phases can also explain the observed ones.

\section{Summary}
In this paper, a three dimensional outer gap model is built to explain the energy dependent light curves of the Vela pulsar observed by $Fermi$ LAT
 (Abdo et al., 2010b). In the present two-layer model,
the acceleration and emission region in the gap  consist two parts, that is,
 the main acceleration region and the screening region.
In the main acceleration region, the charge density is much lower than
the Goldreich-Julian charge density  and
 a strong electric field accelerates the particles to emit GeV photons
via the curvature radiation process. In  the screening region, the charge density is higher than the Goldreich-Julian charge density to screen out the accelerating electric field. We extend our two-dimensional two-layer model in Wang et al. (2010) into a three-dimensional one with the azimuthal structure of the
fractional gap thickness ($f$),
ratio of the thicknesses of the primary and whole region ($h_1/h_2$),
and the number density in the main acceleration region $(1-g_1)$.  Using constant $1-g_1$, $h_1/h_2$ and $f$,
 although the calculated light curves can qualitatively explain
 the observations of the Vela pulsar, the energy dependent light curves
 can not have the third peak. Therefore,
 we considered the  possible azimuthal distributions of these three parameters.
 We found that the distributions of $1-g_1$ and $f$ make third-peak-like
structure in the bridge region of light curve of above  1~GeV,
while the distribution of $h_1/h_2$ makes a bump in the bridge region of
the light curves below 1~GeV.  {{}We find that
the phases of the third peaks caused by the azimuthal
distributions of $h_1/h_2$, $1-g_1$ and $f$ are different from each other.
Consequently, the difference in the phases produces
the shift of the combined third peak with the photon energy.} We also showed that
the present model can reproduce the distribution of the
 cut-off energy for each rotation phase.

\section*{Acknowledgement}
We thank L. Zhang for useful discussion and an anonymous referee for his useful suggestions. This work is supported by a GRF grant of the the Hong Kong SAR Government entitled ``Gamma-ray Pulsars''.

\end{document}